\documentclass{80SA}
\usepackage{times}
\usepackage{color}

\title{Electron Mass Enhancement due to Anharmonic Local Phonons}

\author{Kunihiro Oshiba and Takashi Hotta}

\inst{Department of Physics, Tokyo Metropolitan University,
Hachioji, Tokyo 192-0397, Japan}

\abst{
In order to understand how electron effective mass is enhanced
by anharmonic local oscillation of an atom in a cage composed of
other atoms, i.e., {\it rattling},
we analyze anharmonic Holstein model by using a Green's function method.
Due to the evaluation of an electron mass enhancement factor $Z$,
we find that $Z$ becomes maximum
when zero-point energy is comparable with potential height
at which the amplitude of oscillation is rapidly enlarged.
Cooperation of such quantum and rattling effects is
considered to be a key issue to explain the electron mass enhancement
in electron-rattling systems.
}

\kword{Heavy electrons, Rattling, Sm-based filled skutterudite}

\begin{document}

\maketitle


In Sm-based filled skutterudite SmOs$_{4}$Sb$_{12}$,\cite{Sanada}
it has been observed that electronic specific heat coefficient
$\gamma_{\rm e}$ is significantly enhanced,
suggesting the emergence of heavy-electron state.
An intriguing issue is that $\gamma_{\rm e}$ is almost unchanged
even when we apply a magnetic field up to 30 Tesla.
Namely, the heavy-electron state is magnetically robust.
On the basis of the standard Kondo effect originating
from spin degree of freedom,
it is difficult to explain that heavy effective mass is independent
of the applied magnetic field.
Thus, the non-magnetic Kondo effect due to electron-phonon interaction
has been pointed out,\cite{Miyake}
since local anharmonic oscillation of rare-earth atom,
i.e., {\it rattling}, in the pnictogen cage has played a crucial role
for electronic properties of filled skutterudites.

On such a background, Kondo effect with phonon origin and
its related phenomena have been vigorously investigated.
As an extension of the two-level Kondo problem,
the four- and six-level Kondo systems have been analyzed
to understand the effect of rattling in filled skutterudites.
\cite{Hattori1,Hattori2}
The origin of heavy mass has been discussed
in periodic Anderson-Holstein model~\cite{Mitsumoto}
and in Holstein model.\cite{Fuse}
Due to the evaluation of $\gamma_e$ under the magnetic field
in Anderson-Holstein model,\cite{Hotta1}
it has been found that large $\gamma_{\rm e}$ becomes
magnetically robust,
when the bottom of the potential is relatively wide and flat.
From these efforts, it has been gradually recognized that
magnetically robust heavy-fermion phenomenon actually occurs
in electron-rattling systems,\cite{Hotta2}
but it is further necessary to clarify properties of the heavy-electron
state in electron-rattling systems from various aspects.
In particular, the analysis in the periodic system is required.

In this paper, we analyze anharmonic Holstein model
by using the standard Green's function method,
in order to understand the mechanism of electron mass enhancement
due to rattling.
For the purpose, we evaluate an electron mass enhancement factor $Z$
by focusing on the relation between zero-point energy $E_{0}$
and a potential height $V_0$.
Note that $V_0$ is the potential energy at which the amplitude
of oscillation is rapidly increased.
For $V_{0} > E_{0}$, $Z$ exhibits a peak at $T \sim V_0$,
where $T$ is a temperature,
indicating that the amplitude of oscillation is relevant to
the enhancement of $Z$.
For $V_0 < E_0$, we do not find the peak in $Z$,
but the shoulder structure is observed at $T \sim E_0$.
When we change the states between the cases of
$V_{0} > E_{0}$ and $V_0 < E_0$,
it is found that $Z$ takes the maximum value for $V_0 \sim E_0$.
Thus, we envisage a scenario that $Z$ is significantly enhanced
when both quantum and rattling effects work cooperatively.


Let us introduce anharmonic Holstein model, given by
\begin{equation}
   H = \sum_{\mib{k},\sigma} \varepsilon_{\mib k}
   c_{\mib{k}\sigma}^{\dag} c_{\mib{k}\sigma}
   +g \sum_{\mib{i},\sigma}
     c_{\mib{i}\sigma}^{\dag} c_{\mib{i}\sigma}Q_{\mib{i}}
   +H_{\rm ph},
\end{equation}
where $c_{\mib{k}\sigma}$ indicates an annihilation operator of electron
with momentum $\mib{k}$ and spin $\sigma$,
$\varepsilon_{\mib{k}}$ denotes electron energy,
$g$ is an electron-phonon coupling constant,
$\mib{i}$ denotes a site,
$Q_{\mib{i}}$ is normal coordinate of oscillation of
an atom at site $\mib{i}$,
and $c_{\mib{i}\sigma}$ is an annihilation operator of electron
at site $\mib{i}$.
Throughout this paper, we use such units as $\hbar$=$k_{\rm B}$=1.

The third term in eq.~(1) indicates the oscillation of atom, given by
$H_{\rm ph}$=$\sum_{\mib{i}} [P^2_{\mib{i}}/2+V(Q_{\mib{i}})]$,
where $P_{\mib{i}}$ is the canonical momentum and
$V(Q_{\mib{i}})$ is an anharmonic potential for atom.
Note that the reduced mass of the oscillation is set as unity.
The potential is expressed by
$V(Q_{\mib{i}})$=$\omega^2 Q_{\mib{i}}^{2}/2$+$k_{4} Q_{\mib{i}}^{4}$+
$k_{6} Q_{\mib{i}}^{6}$,
where $\omega$ is the energy of oscillator and
$k_{4}$ and $k_{6}$ denote the coefficients
for forth- and sixth-order anharmonic terms, respectively.
Here we introduce the phonon operator $a_{\mib{i}}$ defined through
$Q_{\mib{i}}$=$(a_{\mib{i}}+a_{\mib{i}}^{\dagger})/\sqrt{2\omega}$.
Then, we obtain
\begin{equation}
  H_{\rm ph} \!=\! \omega \sum_{\mib{i}}
   [a_{\mib{i}}^{\dagger} a_{\mib{i}} \!+\! 1/2
   \!+\! \beta (a_{\mib{i}} + a_{\mib{i}}^{\dagger})^{4}
   \!+\! \gamma (a_{\mib{i}} + a_{\mib{i}}^{\dagger})^{6}],
\end{equation}
where $\beta$ and $\gamma$ are non-dimensional coefficients of
anharmonic terms,
given by $\beta$=$k_{4}/(4\omega^{3})$ and $\gamma$=$k_{6}/(8\omega^{4})$.
By using such non-dimensional parameters, we express the potential as
$V(q_{\mib{i}})$=$\alpha \omega (q_{\mib{i}}^{2}+16 \alpha \beta q_{\mib{i}}^{4}+64 \alpha^{2} \gamma q_{\mib{i}}^{6})$,
where $\alpha$ is the non-dimensional electron-phonon coupling constant
given by $\alpha$=$g^{2}/(2\omega^{3})$ and $q_{\mib{i}}$=$Q_{\mib{i}}/\ell$.
Here $\ell$ denotes the typical length for oscillation,
given by $\ell$=$g/\omega^2$.


First let us briefly discuss
how the potential shape is changed by anharmonic parameters.
Note that in this paper, we consider only the case of $\gamma>0$,
since the oscillation should be confined in a finite space.
When we change the value of $\beta$,
the potential shapes are classified into three:\cite{Hotta1}
(i) On-center type for $-\sqrt{3\gamma/4}<\beta<0$,
(ii) rattling type for $-\sqrt{\gamma}<\beta \leq -\sqrt{3\gamma/4}$,
and
(iii) off-center type for $\beta \leq -\sqrt{\gamma}$.
As shown in Fig.~1,
the on-center type potential has a minimum at $q$=0,
while in the off-center type, we observe three minima.
The rattling-type potential has relatively wide region in the bottom.
Since we are interested in the effect of oscillation
in the rattling-type potential,
here we consider the case of $\beta$=$-\sqrt{3\gamma/4}$,
depicted by thick curve in Fig.~1.
In such a case, the potential has two saddle points and
we define $V_0$ as the potential height at the saddle point.
When we increase the energy of oscillator, the amplitude of oscillation
is considered to be rapidly enlarged at the energy of $V_0$.

\begin{figure}
\begin{center}
\includegraphics[width = 85mm,angle = 0]{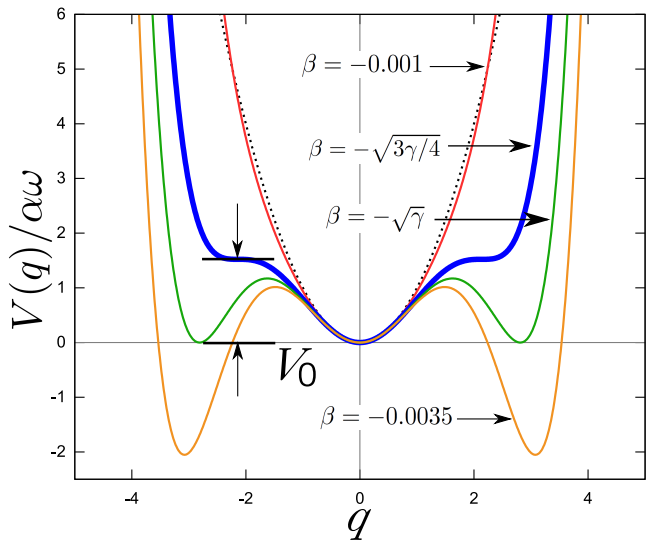}
\end{center}
\caption{(Color online)
Change of the potential shape at $\gamma$=$10^{-5}$.
Dotted curve denotes harmonic potential shape for comparison.
The curve of $\beta$=$-0.001$ denotes the on-center type,
while that of $\beta$=$-0.0035$ indicates the off-center one.
The curves of $\beta$=$-\sqrt{3\gamma/4}$ and $-\sqrt{\gamma}$
are categorized into the rattling type.
In this paper, we focus on the case of $\beta$=$-\sqrt{3\gamma/4}$.
}
\label{potential}
\end{figure}


Now we calculate the electron self-energy $\Sigma$ in the secound-order
perturbation in terms of the coupling $g$ within an adiabatic approximation.
Namely, we consider a situation in which electron bandwidth $W$
is much larger than the phonon energy $\omega$.
Hereafter $W$ is taken as the energy unit and we set $W$=1.
First we evaluate the bare phonon Green's function $D$.
For the purpose, we diagonalize $H_{\rm ph}$
to obtain eigenenergy $E_{n}$ and eigenstates $|{n}\rangle$.
Then, $D$ is obtained in the Lehmann representation as
\begin{equation}
   D(z) \!=\! \frac{1}{\Omega} \sum_{n,m}
   \frac{e^{-E_{n}/T}-e^{-E_{m}/T}}{z+E_{n}-E_{m}}
   |\langle m|(a_{\mib{i}}+a_{\mib{i}}^{\dagger})|n\rangle|^2,
\end{equation}
where $z$ is a complex number with the dimension of energy
and $\Omega$ is a partition function.
Note that the site dependence does not appear in $D$,
since we consider Einstein-type local phonons.
Then, we evaluate the electron self-energy as
\begin{equation}
  \Sigma(i\omega_n) = -\alpha \omega^{2} T \sum_{n'}\sum_{\mib{k}'}
  D(i\omega_{n}-i\omega_{n'})G(\mib{k}',i\omega_{n'}),
\end{equation}
where $G$ denotes the electron Green's function
given by $G(\mib{k},i\omega_{n})$=$1/(i\omega_n-\varepsilon_{\mib{k}})$
and $\omega_{n}$ is fermion Matsubara frequency
defined by $\omega_{n}$=$(2n+1)\pi T$ with an integer $n$.
By assuming the electron density of states with rectangular shape
of the width $W$, we perform the summation in terms of $\mib{k}'$.
Note that the half-filling case is considered in this paper.
Then, we obtain $\Sigma$ as
\begin{equation}
  \Sigma(i\omega_n) \!=\! i\lambda \omega T \sum_{n'}
  D(i\omega_{n}-i\omega_{n'})
  \tan^{-1}\Bigl(\frac{W}{2 \omega_{n'}}\Bigr),
\end{equation}
where $\lambda$ denotes the Eliashberg electron-phonon
coupling constant,
defined by $\lambda$=$2\alpha \omega/W$.
In actual calculations for $\Sigma$,
we exploit a Fast-Fourier-Transformation technique.
Then, we define the mass enhancement factor $Z$ as
\begin{equation}
  Z = 1 - \Sigma(i\pi T)/(i\pi T).
\end{equation}
Here we use the value on the imaginary axis
at the lowest temperature as an approximation for $Z$,
although $Z$ should be defined by the energy differentiation
of $\Sigma$ on the real axis.
Since $\Sigma$ is found to exhibit a peak at the energy
in the order of $\omega$,
we replace the energy differentiation with the energy difference
of the width $T$ for $T < \omega$.
Thus, we adopt the present definition for $Z$
to discuss properties of $Z$ in the region of $T < \omega$.
In fact, characteristic structure of $Z$ is found for $T<\omega$,
which is consistent with the present condition.

\begin{figure}
\begin{center}
\includegraphics[width = 85mm,angle = 0]{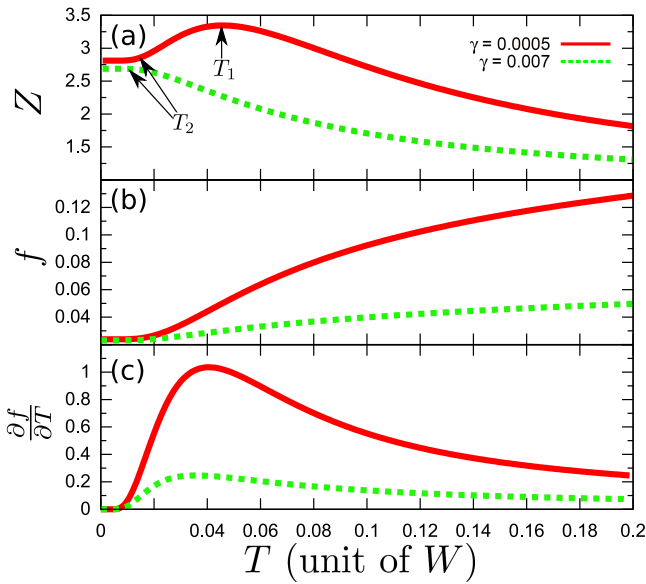}
\end{center}
\caption{(Color online)
(a) Mass enhancement factor $Z$ vs. temperature $T$ for $\gamma$=$0.0005$
with $E_{0} < V_{0}$ and for $\gamma$=$0.007$ with $E_{0} > V_{0}$.
We note that we always keep the relation of $\beta$=$-\sqrt{3\gamma/4}$.
Note also that $T_{1}$ is a peak temperature,
while $T_{2}$ is a temperature at which
$Z$ begins to be constant with decreasing $T$.
(b) Non-dimensional Debye-Waller factor $f$ vs. $T$ and
(c) $\partial f/\partial T$ vs. $T$ for the same parameters as (a).
}
\label{zvst}
\end{figure}


Now let us discuss the calculated results.
Here we set $\lambda$=$1.0$ and $\omega$=0.1.
For the diagonalization of $H_{\rm ph}$, we use 250 phonon basis.
In the calculation of $\Sigma$, we take 4096 Matsubara frequencies.
First we discuss temperature dependence of $Z$
by focusing on the relation between $V_0$ and zero-point energy $E_0$.
In Fig.~2(a), we depict two curves for $\gamma$=$0.0005$ and $0.007$.
Note that the condition of $\beta$=$-\sqrt{3\gamma/4}$ is always kept.
For $\gamma$=$0.0005$, we find that $E_{0}$=$0.00432$ and $V_{0}$=$0.0108$,
suggesting the case of $V_0 > E_0$.
On the other hand, for $\gamma$=$0.007$,
we obtain $E_{0}$=$0.00359$ and $V_{0}$=$0.00288$,
leading to the situation with $V_0 < E_0$.

As observed in Fig.~2(a), for the case of $\gamma$=$0.0005$,
$Z$ has a broad peak at a temperature
which is defined as $T_1$ in this paper.
From the detailed calculations,
it is found that $T_1$ is scaled by $V_0$.
When we further decrease the temperature, we observe the structure of
a shoulder.
We define $T_2$ as the temperature at which $Z$ becomes constant.
For the case of $\gamma$=$0.007$, on the other hand,
we find no peak and only $T_2$ can be defined.
After performing the calculations for several parameter sets,
we have found that $T_2$ is scaled by $E_0$. It is also found that the first-order excitation energy is always
in the order of $E_{0}$ in the present parameters,
suggesting that the temparature scale is considered to be
$E_{0}$ at low temparatures.

From these results, first we claim that the enhancement of $Z$ is
closely related to the potential height $V_0$.
By taking into account the definition of $V_0$, we point out that
$Z$ becomes large when the amplitude of oscillation is rapidly increased.
This should be called the rattling effect.
The result is consistent with the previous conclusion from the numerical
calculation of $\gamma_{\rm e}$ in Holstein-Anderson model.\cite{Hotta1}
Second we claim that the rattling effect is not significant for $T < E_0$,
since zero-point oscillation is dominant in such a region
and the rattling effect is masked.

In order to examine the relation between $Z$ and the amplitude of oscillation,
we evaluate the Debye-Waller factor
which represents the intensity of thermal motion of the caged atom.
In the non-dimensional form, it is given by
\begin{eqnarray}
 f = \frac{F_{\rm DW}}{|\mib{G}|^{2} \ell^2 }
   = \frac{\langle Q^{2}_{\mib{i}} \rangle}{3 \ell^2},
\end{eqnarray}
where $F_{\rm DW}$ denotes the Debye-Waller factor,
$\mib{G}$ denotes the reciprocal lattice vector,
and $\langle \cdots \rangle$ indicates the operation to take thermal average.
Note again that the site dependence is ignored,
since we consider Einstein-type local phonons.

In Fig.~2(b), we show calculated results for $f$.
In order to clarify the variation of $f$ in terms of $T$,
we also plot $\partial f/\partial T$ vs. $T$ in Fig.~2(c).
From these results, we claim that
$f$ rapidly increases at $T \simeq T_{1}$
for the case of $V_0 > E_0$.
If we further decrease $T$, we arrive at a constant value
at $T_2 \sim E_0$.
For the case of $V_0 < E_0$, however,
the value of $\partial f/\partial T$ is not large in comparison
with the case of $V_0 > E_0$,
suggesting that the amplitude of oscillation does not change
so rapidly.
Note that $f$ is constant in the region of $T \leq T_{2}$
both for $V_0 > E_0$ and $V_0 < E_0$.

\begin{figure}
\begin{center}
\includegraphics[width = 85mm,angle = 0]{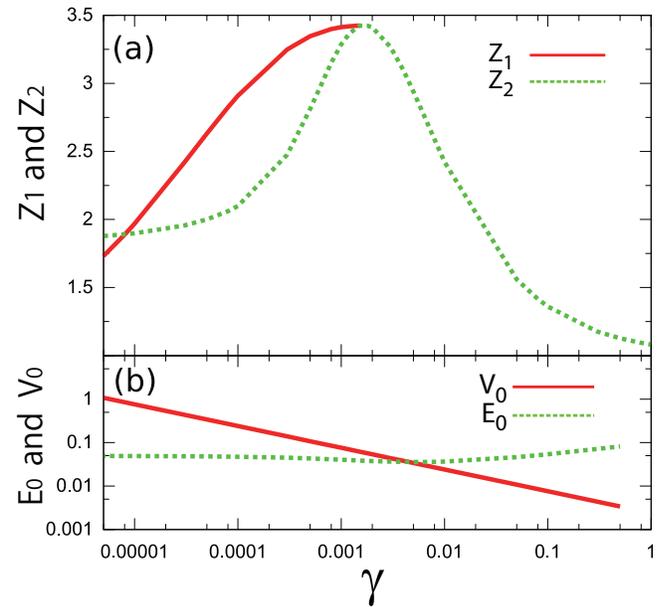}
\end{center}
\caption{(Color online)
(a) $Z_1$ and $Z_2$ vs. $\gamma$ and
(b) $E_{0}$ and $V_{0}$ vs. $\gamma$
for $\beta$=$-\sqrt{3\gamma/4}$.
}
\label{zmax}
\end{figure}

For $V_0 > E_0$, the atom feels the wide potential
near the saddle point and
it can oscillate with large amplitude at $T_{1} \sim V_{0}$.
Since $Z$ exhibits the peak around at $T$=$T_{1}$,
we conclude that the enhancement of $Z$ is related with
the rattling effect, which is caused by
the rapid change of the amplitude of oscillation.
For $V_0 < E_0$, on the other hand,
since the zero-point energy of the atom is larger than $V_{0}$,
the rattling effect is masked by quantum fluctuations
in the low-temperature region of $T < E_{0}$.
Thus, we arrive at the saturated value of $Z$
dominated by zero-point oscillation at $T \sim T_2$.
In this case, $Z$ does not exhibit a peak in the temperature dependence.

Next we consider how $Z$ is changed by $\gamma$ at $T$=$T_{1}$ and $T_{2}$.
For the purpose, here we define $Z_1$ and $Z_2$ as
the values of eq.~(6) at $T$=$T_1$ and $T_2$, respectively.
In Fig.~3(a), we depict $Z_1$ and $Z_2$ vs. $\gamma$.
For large value of $\gamma$, we find only the shoulder stricture
characterized by $T_2$, as observed in the curve for $\gamma$=$0.007$
in Fig.~2(a).
The value of $Z_2$ is increased with the decrease of $\gamma$
and eventually, it turns to be decreased to form a peak
at a certain value of $\gamma$, which is defined as $\gamma^*$.
At $\gamma$=$\gamma^*$, we also find $T_1$ in addition to $T_2$,
as observed in the curve for $\gamma$=$0.0005$ in Fig.~2(a).
When we further decrease $\gamma$ less than $\gamma^*$,
both $Z_1$ and $Z_2$ are decreased.
At very small $\gamma$, $Z_1$ becomes smaller than $Z_2$,
suggesting the complicated multi-peak structure
in the temperature dependence in $Z$,
but we do not show such a result,
since it is out of the scope of this paper.

Here we focus on the peak of $Z$ at $\gamma$=$\gamma^*$.
In order to clarify the meaning of $\gamma^*$,
we plot $E_{0}$ and $V_{0}$ as functions of $\gamma$ in Fig.~3(b).
Except for small deviation, we find that $\gamma^*$
indicates the value of $\gamma$ at which $E_0$ and $V_0$
cross with each other.
As we have emphasized above, for the case of $V_0 > E_0$,
we find the peak in $Z$ at $T \sim V_0$ due to the rattling effect.
For the case of $V_0 < E_0$, the ratting effect is
masked by quantum fluctuations and we do not find the peak
induced by rattling effect.
At $V_0 \sim E_0$, it is expected that both quantum and rattling effects
work cooperatively for the enhancement of $Z$.
Thus, $Z$ is considered to take the maximum for $V_0 \sim E_0$.


In summary, we have discussed the electron mass enhancement factor $Z$
in anharmonic Holstein model.
It has been found that $Z$ is enhanced by the rattling effect,
i.e., the rapid change of the amplitude of oscillation.
Furthermore, $Z$ has been found to take the maximum value
when $E_{0}$ is comparable with $V_{0}$.
Namely, the heavy-electron state due to rattling is expected
to occur significantly when zero-point energy is comparable with
the energy at which the amplitude of oscillation is rapidly
increased.


This work has been supported by a Grant-in-Aid for Scientific Research
on Innovative Areas ``Heavy Electrons''
(No. 20102008) of The Ministry of Education, Culture, Sports,
Science, and Technology, Japan.


\end{document}